\documentclass{emulateapj}

\pdfoutput=1

\usepackage{tmsfnts}

\shorttitle{\emph{INTEGRAL} OBSERVATIONS OF THE COMA CLUSTER}

\shortauthors{LUTOVINOV ET AL.}

\begin{document}

\title{X-ray observations of the Coma Cluster in a broad energy band with
\emph{INTEGRAL}, \emph{RXTE} and \emph{ROSAT} observatories.}

\author{A.~A.~Lutovinov\altaffilmark{1,2},
  A.~Vikhlinin\altaffilmark{2,1}, E.~M.~Churazov\altaffilmark{3,1},
  M.~G.~Revnivtsev\altaffilmark{3,1}, R.~A.~Sunyaev\altaffilmark{3,1}}

\slugcomment{Submitted to ApJ 2/24/2008; Revised 5/15/08; arXiv:0802.3742}

\altaffiltext{1}{Space Research Institute, Moscow,
Russia}\email{aal@hea.iki.rssi.ru}

\altaffiltext{2}{Harvard-Smithsonian Center for Astrophysics, Cambridge,
  02138 MA, USA} \altaffiltext{3}{Max-Planck-Institut f\"ur Astrophysik, Garching,
  Germany}

\begin{abstract}
  We present results of X-ray observations of the Coma cluster with multiple
  instruments over a broad energy band. Using the data from \emph{INTEGRAL},
  \emph{RXTE} and \emph{ROSAT} observatories, we find that the Coma spectrum
  in the $0.5-107$ keV energy band can be well approximated by a thermal
  plasma emission model with a temperature of $T=8.2$ keV. \emph{INTEGRAL}
  was used to image the cluster emission in the hard energy band. The
  cluster is only marginally detectable ($\sim1.6\,\sigma$) in the
  $44-107$~keV energy band; however, the raw flux in this band is consistent
  with the previous results from \emph{Beppo-SAX} and \emph{RXTE}
  observatories. We can exclude with high significance that the hard-band
  flux reported by \emph{Beppo-SAX} and \emph{RXTE} could be produced by a
  single point source. The $20-80$ keV flux of a possible non-thermal
  component in the cluster spectrum is $(6.0\pm8.8)\times10^{-12}$ ergs
  cm$^{-2}$ s$^{-1}$. It is unlikely that the IC scattering of CMB photons
  is able to produce hard X-ray flux at these levels, unless the magnetic
  field strength is as low as 0.2 $\mu$G. The latter value can be considered
  as a lower limit on the field strength in Coma. We also present a
  temperature map of the central part of the cluster, which shows
  significant variations and in particular, a hot, $\sim11.5$ keV, region in
  the extension towards the subcluster infalling from the South-West.
\end{abstract}



\section{Introduction}

Hot intracluster gas should have an admixture of non-thermal components,
including relativistic electrons, magnetic fields, relativistic protons, and
supra-thermal electrons. Some of these components are expected on the
theoretical grounds while the presence of relativistic electrons and sizable
magnetic fields is clear from observations of cluster radio halos \citep[see
e.g., a recent review by][]{fer08}.  Often, the very existence of
non-thermal components in the ICM poses interesting problems which have been
studied intensively \citep[see][for a recent review]{2008SSRv..tmp...16R}.
The relativistic electron population in clusters can be studied in the hard
X-ray band through the high-energy spectral components observed on top of
the thermal bremsstrahlung spectra. If such components are present at the
intensity levels accessible with the current instrumentation, their likely
origin is inverse Compton (IC) scattering of the cosmic microwave background
photons; the models in which high-energy X-rays are produced by
bremsstrahlung of the supra-thermal electrons are less attractive (e.g.,
Petrosian, 2001; Petrosian, Bykov \& Rephaeli, 2008; Petrosian \& Eats,
2008) because subrelativistic electrons are subject to strong Coulomb losses
(but see \cite{dog07}). At some level, an excess around 20--30~keV with
respect to a single-temperature fit can be attributed to the presence of
localized patches of hot gas with temperatures substantially exceeding the
cluster average. The amplitude and the shape of such high energy excess is
an important diagnostic for the non-isothermality of the gas.

A classic application of the hard X-ray spectral observations is the
measurement of the bulk magnetic field strength via comparison of the
inverse Compton and radio synchrotron emission (e.g., Felten \& Morrison
1966, Tucker et al., 1973, Rephaeli 1979). A prime object for such studies
is the Coma cluster, which possesses a bright, well-studied radio halo (see,
e.g., \citet{thi03}).  Since Coma is also the nearest rich cluster, it was
observed with virtually every X-ray observatory flown.

Detection of the hard X-ray component in excess over the extrapolation of
the thermal spectrum in Coma was first reported from the \emph{RXTE} and
\emph{Beppo-SAX} observations \citep{rep99,fus99}, later confirmed by more
extensive analyses \citep{rep02,fus04}. The statistical significance of
these detections remains not very high \citep[e.g.,][reported a
$\sim4.8\sigma$ detection]{fus04}, and the reported detections are subject
to criticism \citep[e.g.,][report only a $\sim 2\sigma$ significance of the
hard component from independent reanalysis of the data]{ros04}.  Even if a
hard component exists in the Coma spectrum, we cannot exclude, on the basis
of the \emph{Beppo-SAX} of \emph{RXTE} data, a possibility that it is
produced by, e.g., a strongly absorbed AGN.  Obviously, the situation can be
improved through observations of Coma with an \emph{imaging} hard X-ray
telescope.

Telescopes of the \emph{INTEGRAL} observatory \citep{win03} offer a unique
combination of good sensitivity and angular resolution in hard X-rays. The
total \emph{INTEGRAL} exposure of the Coma cluster is almost $10^6$~sec,
leading to the sensitivity for an extended source at $E \sim 50$~keV which
is comparable to that of the \emph{Beppo-SAX} and \emph{RXTE} data. Coma is
detected by \emph{INTEGRAL} with high significance at $E<30$~keV, where the
emission is dominated by the thermal component; the analysis in this energy
band has been reported by \citet{ren06} and \citet{eck07}; \citet{ren06}
also used the first 500~ksec of the Coma data to put limits on the hard
X-ray component. In this work, we present a systematic analysis of the hard
X-ray component using the full \emph{INTEGRAL} exposure in combination with
the data from \emph{ROSAT} and \emph{RXTE}. The combination of the data from
these satellites provides an accurate, self-consistent measurement of the
broad band thermal spectrum of Coma, which is critical for detection of
non-thermal components at or below 50~keV. We also present a temperature map
of the central region of Coma obtained from the ratio of \emph{INTEGRAL} and
\emph{ROSAT} brightnesses.

All distance-dependent quantities are reported assuming that Coma is at
$d=90.5$~Mpc (corresponding to $z=0.023$ and $H_0=70$ km s$^{-1}$
Mpc$^{-1}$).

%\newpage

\section{Observations and data analysis}

We use X-ray observations of the Coma cluster performed by \emph{ROSAT},
\emph{RXTE}, and \emph{INTEGRAL}. These observatories combined cover a very
broad energy range, 0.5--100~keV, and thus provide a very high-quality broad
band spectrum. Since Coma is a bright X-ray source, the spectral
measurements with each instrument are relatively straightforward. The main
complication is adjusting the observed spectra to account for the different
spatial response: \emph{ROSAT} is a direct-imaging telescope;
\emph{INTEGRAL} is a coded-mask imager which results in a rather broad Point
Spread Function (PSF) in reconstructed images; \emph{RXTE} is a collimator
with a beam pattern which is slightly smaller than the cluster size. Also,
the statistical quality of the Coma data in the \emph{ROSAT} and \emph{RXTE}
energy bands is very high and so uncertainties are dominated by the
instrument cross-calibration. These issues are discussed below for each
telescope individually.

\subsection{\emph{INTEGRAL}}

Coma was targeted by \emph{INTEGRAL} in several sets of observations with a
total exposure near $106$~seconds (revolutions 36, 71, 72, 274, 275, 317,
318, 319, 324, 325). We concentrated here on the data from ISGRI detector
\citep{leb03} of the IBIS telescope \citep{ube03}. This telescope provides a
wide field of view ($28\degr\times28\degr$) and an angular resolution of
12\arcmin. The telescope is sensitive over approximately $17-200$~keV energy
band. Some fraction of the IBIS data was affected by Solar flares and was
discarded from the analysis. The clean set of IBIS data was initially
reduced in individual pointings (so called science windows), each with a
typical duration $\sim 2$~ksec. The reconstructed images were combined into a
single mosaic which was analyzed further. The total dead-time corrected
combined exposure was $990$ ksec.

The IBIS/ISGRI image reconstruction algorithm was discussed previously in
\citet{kri05,kri07}, and we refer the reader to these works. A crucial
prerequisite for this --- and any other coded mask image reconstruction
algorithm --- is the ability to accurately predict the detector background
image in the absence of any sources. Our image reconstruction software uses
standard calibration tables (OSA 6.0\footnote{http://isdc.unige.ch}) were
used to correct the event energies depending of their rise-time. We applied,
however, additional corrections associated with the secular change of gain
\citep[see discussion in][]{tsy06}.

%%%%%%%%%%%%%%%%%%%%%%%%%%%%%%%%%%%%%%%%%%%%%%%%%%%%%%%%%%%%%%%%%%%%%%%
\begin{figure}
\centerline{\includegraphics[width=\columnwidth,bb=1 2 560 500,clip]{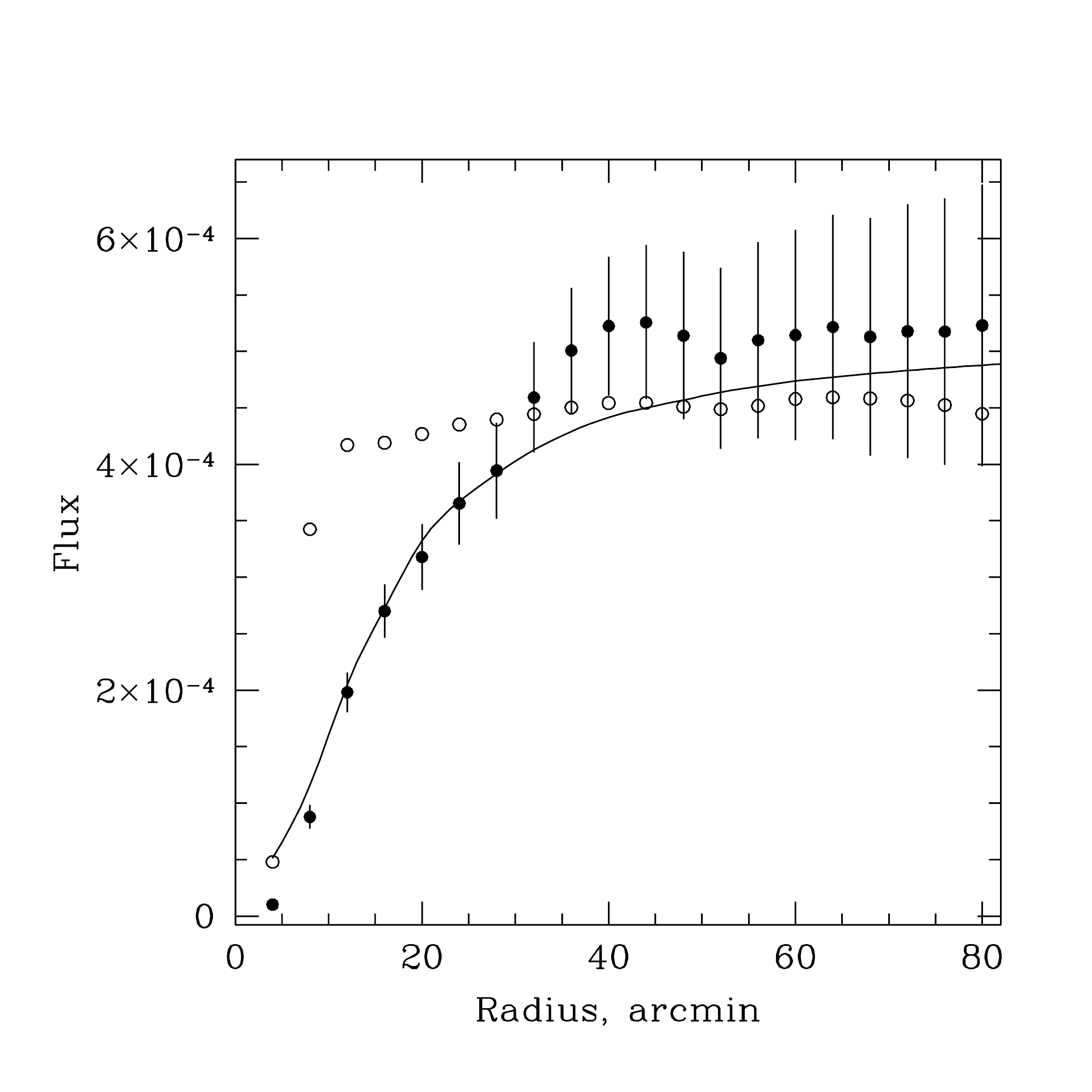}}
\caption{The ``growth functions'' (flux integrated within a given radius in
  the reconstructed image) of the Coma cluster (solid circles) and Crab
  nebula (open circles) in the $17-22$ keV energy band. The solid line shows
  a growth curve expected for a $\beta$-model profile with $\beta=0.741$ and
  $r_c=10.7$\arcmin{} (parameters for the $\beta$-model in the \emph{ROSAT}
  band). The Crab flux was rescaled to the Coma level for comparison.}
\label{growth}
\end{figure}
%%%%%%%%%%%%%%%%%%%%%%%%%%%%%%%%%%%%%%%%%%%%%%%%%%%%%%%%%%%%%%%%%%%%%%

Angular size of the Coma cluster is significantly larger than the IBIS PSF.
This is apparent from the comparison of the ``growth function'' (flux
integrated within an aperture) for Coma and a point source (Crab) shown in
Fig.\ref{growth}. While essentially all of Crab flux is concentrated within
12\arcmin{} (three 4\arcmin{} pixels in the reconstructed image), for Coma
the curve continues to grow up to a 30\arcmin{} radius. In fact, the
observed growth curve is consistent with that for a $\beta$-model (Cavaliere
\& Fusco-Femiano 1976) fit to the \emph{ROSAT} image, $S(r)=S_0
(1+r2/r_c2)^{-3\beta+0.5}$, with $\beta=0.741$ and $r_c=10.7$\arcmin.
Because of such a good agreement, the total Coma flux in the \emph{INTEGRAL}
energy bands can be obtained by fitting a normalization of the $\beta$-model
instead of directly integrating the flux in each pixel within a wide
aperture. The advantage of this method is that it provides a higher
statistical accuracy since each pixel is added with an optimal weight.
Direct integration, however, is less model-dependent since the hard-band
thermal --- and especially, the inverse Compton --- emission in general does
not have to follow the distribution of the surface brightness in the
\emph{ROSAT} band.  Given the pros and cons, our choice is to use the
$\beta$-model fluxes throughout but always check that they are consistent
with the direct integration.

The limiting factor for the \emph{INTEGRAL} observations of the Coma cluster
is the level of background fluctuations. In addition to purely statistical
fluctuations, the background can leave systematic variations in the
reconstructed image if subtracted incompletely from the raw detector image.
Every care was taken, therefore, to ensure that the background was
subtracted correctly. Background templates were taken from observations of
``empty'' fields closest in time to each Coma pointing or group of
pointings. With this approach, we correctly take into account all possible
temporal variations of ISGRI background. As a final check, we checked the
statistics of fluctuations in the reconstructed image (similar to the
analysis in \S 3 of \citet{kri07}), and also fluxes within 40\arcmin{}
circles at different off-axis locations. In both cases, the variations had
the mean consistent with zero and the dispersion equal to that expected for
the purely Poisson fluctuations. We conclude, that the accuracy of our
background modeling has reached its fundamental statistical limit (for the
exposure time accumulated during Coma observations).

To convert ISGRI counts to physical flux units (erg~s$^{-1}$~cm$^{-2}$), we
used the calibration observations of the Crab nebula which are regularly
performing by \emph{INTEGRAL}. Crab observations were reduced with the same
software setup we used for Coma and the counts-to-flux conversion
coefficients were determined assuming the ``conventional'' spectral
parameters for Crab, $I=9.7$~ phot~s$^{-1}$~cm$^{-2}$~keV$^{-1}$
\citep{too94}.

After experimenting with energy bands for the Coma analysis, we chose to
extract fluxes in the $17-22$, $22-28.5$, $28.5-44$, and $44-107$~keV
bands.  ISGRI efficiency quickly drops below 17~keV setting a natural
lower limit for the energy band. The width of the two lower channels,
factor of 1.3 in energy, was chosen to ensure high statistical
significance of the Coma detection ($>7\sigma$). The flux in the
$44-107$~keV channel should be dominated by non-thermal components. The
third channel, $28.5-44$~keV fills the gap.

Source fluxes were extracted within the 60\arcmin\ radius. This aperture,
one the one hand, contains almost all of the \emph{INTEGRAL} X-ray flux
(Fig.\ref{growth}), and on the other hand matches the size of the \emph{RXTE} field
of view, making flux comparisons easier. \emph{ROSAT} pointings fully cover
this region. 

%%%%%%%%%%%%%%%%%%%%%%%%%%%%%%%%%%%%%%%%%%%%%%%%%%%%%%%%%%%%%%%%%%%%%
\begin{figure}
\vspace*{-3mm}
\centerline{\includegraphics[width=\columnwidth]{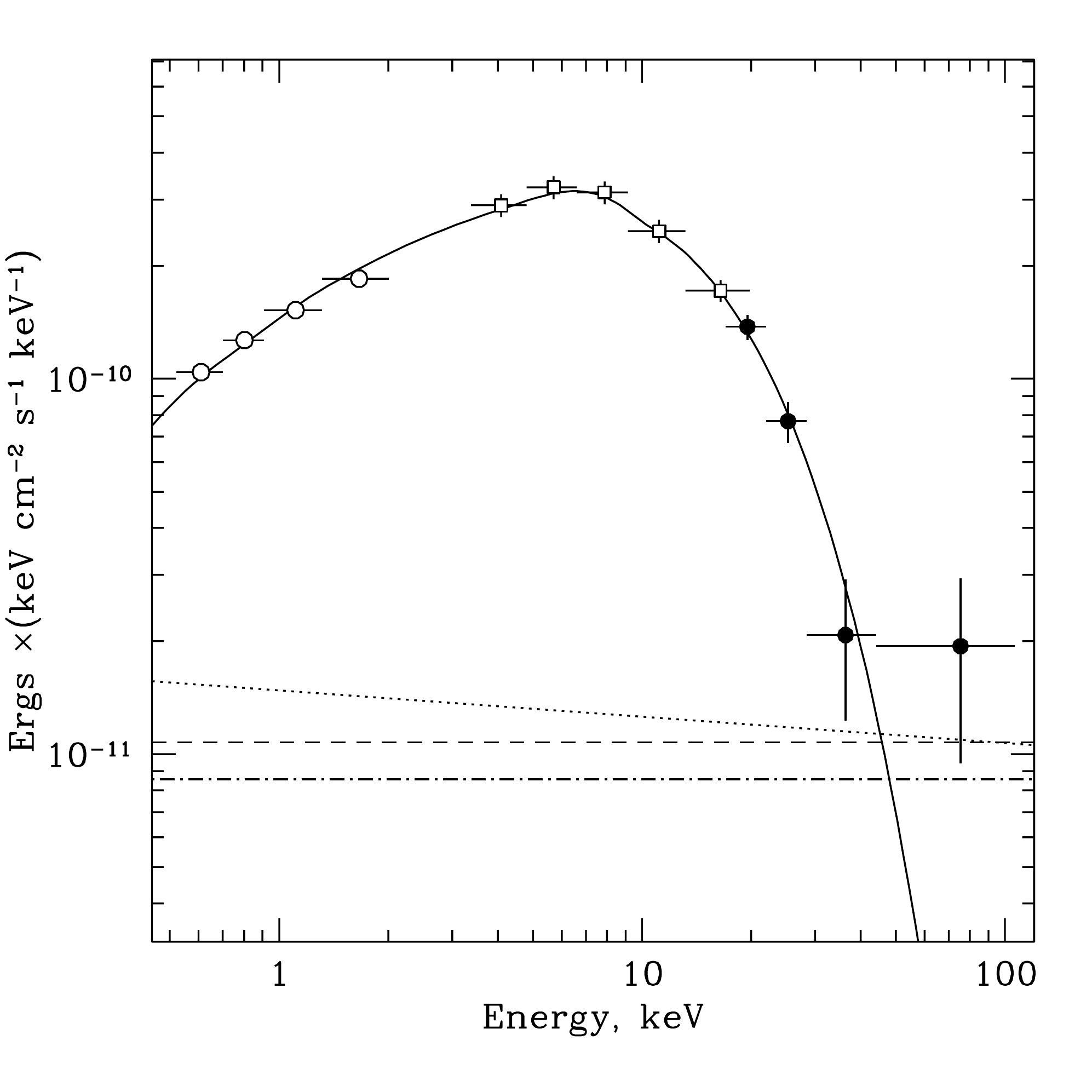}}
\caption{Broadband spectrum of the Coma cluster from \emph{INTEGRAL} (filled
  circles), \emph{RXTE} (open squares), and \emph{ROSAT} (open circles). The
  best-fit MEKAL model is shown by solid line. The contribution of
  non-thermal components corresponding to the \citet{rep02} and
  \citet{fus04} detections is shown by dotted and dashed lines,
  respectively. A $2\sigma$ upper limit for a point source emission is shown
  by a dashed-dotted line.}
  \label{spectrum}
\end{figure}
%%%%%%%%%%%%%%%%%%%%%%%%%%%%%%%%%%%%%%%%%%%%%%%%%%%%%%%%%%%%%%%%%%%%%

\subsection{\emph{RXTE}}

Coma was observed by \emph{RXTE} for $\sim90$~ksec (Obs.ID 10368) and
the data from its PCA spectrometer can be used to measure the cluster
spectrum in the $3-20$ keV energy band.  The data reduction was done
with standard programs of the LHEASOFT package (v6.0). To increase the
sensitivity and reduce systematic uncertainties, we used only data of
the first layers of PCA detectors. The background was based on the
``L7\_240CM'' model (this includes both the particle-induced detector
background and the all-sky average of the Cosmic X-ray Background).

One of the complications in the \emph{RXTE} analysis is to correctly compute
the flux fraction outside the PCA collimator FOV. The PCA beam pattern can
be modeled as a convolution of a near-conical response of the individual
collimator, $R(x,y)$, with a Gaussian ($\sigma=6'$) which corresponds to
misalignment of individual collimators and spacecraft pointing drift during
the observation \citep{jah06}. $R(x,y)$ is tabulated in the calibration file
{\it pcacol}. Assuming that in the PCA energy band, the Coma surface
brightness distribution follows the \emph{ROSAT} image, the effective flux
fraction within the PCA FOV can be computed as
\begin{eqnarray}
  \frac{\int S(x,y) \times (R(x,y)\otimes G(x,y))\, dx dy}
  {\int S(x,y)\,dx dy} = 0.763,
\end{eqnarray}
where $S(x,y)$ is the \emph{ROSAT} image and $G(x,y)$ is a Gaussian. To
obtain total Coma fluxes, the observed \emph{RXTE} count rates should be
divided by this factor.

In addition to the FOV correction, we need to ensure good cross-calibration
of the \emph{RXTE}/PCA with \emph{ROSAT} and \emph{INTEGRAL}. To this end,
we note that the best-fit parameters of the Crab spectrum obtained with the
standard PCA response matrix differ significantly from the
``conventional'' values: $2.09\pm0.04$ and $11.6\pm0.4$ phot s$^{-1}$
cm$^{-2}$ keV$^{-1}$ instead of 2.1 and 9.7 phot s$^{-1}$ cm$^{-2}$
keV$^{-1}$, respectively, indicating a possible problem with the PCA
absolute calibration \citep{rev03}. An ad-hoc correction factor,
\begin{eqnarray}
f_{corr}(E)=0.836 E^{-0.01}
\end{eqnarray}
applied to the observed PCA spectra brings the Crab results into agreement
with the conventional average spectrum \citep{too94}. We apply this
correction factor also to the Coma data. We note that this eliminates a 20\%
difference in flux when the PCA spectrum is extrapolated to the \emph{ROSAT}
band; the \emph{INTEGRAL} and PCA fluxes near 20~keV are also in a good
agreement after this correction is applied (Fig.\ref{spectrum}). Since we
effectively base the PCA response calibration on the absolute Crab spectrum,
we need to assign systematic errors to account for the uncertainties in the
latter. The uncertainty of the Crab spectral index, $\pm0.04$ \citep{rev03}
can be approximated if we bin the PCA data into five wide energy channels
and assign a 7\% uncertainty to the flux in each channel.

\subsection {\emph{ROSAT}}

The \emph{ROSAT} PSPC pointed observations of Coma were reduced as described
in \citet{vih99}. The reduction pipeline was based on S.~Snowden's software
\citep{sno94}. This software eliminates periods of high particle and
scattered solar backgrounds as well as those intervals when the detector may
be unstable. Exposure maps in several energy bands are then created using
detector maps obtained during the \emph{ROSAT} All-Sky Survey.  The exposure
maps include vignetting and all detector artifacts. The unvignetted particle
background is estimated and subtracted from the data even though the PSPC
particle background is low compared to the cosmic X-ray background.  The
scattered solar X-ray background also should be subtracted separately,
because, depending on the viewing angle, it can introduce a constant
background gradient across the image.  Most of Solar X-rays were eliminated
by simply excluding time intervals when this emission was high, but the
remaining contribution was also modeled and subtracted. The remaining
background was estimated by extracting the radial surface brightness profile
from the merged data, and fitting it to the $\beta+\mathrm{const}$ model at
large radii. The background-subtracted images are suitable for direct
extraction of fluxes in the energy bands of interest. Our imaging analysis
uses the 0.5--2~keV band images.

\section{Results}

\subsection{Spectral analysis}

The combined energy spectrum of Coma from \emph{INTEGRAL} ($17-107$~keV),
\emph{RXTE} ($3-20$~keV), and \emph{ROSAT} ($0.5-2$~keV) is shown in
Fig.\ref{spectrum}. The spectrum can be well fit with the thermal plasma
emission (MEKAL) with the temperature $T=8.2\pm0.2$~keV, abundance fixed at
$0.250$ Solar \citep{arn01}, and Galactic absorption
$N_H=9\times10^{19}$~cm$^{-2}$. The best-fit approximation is shown in
Fig.\ref{spectrum} by a solid line. This model gives a flux in the
\emph{INTEGRAL} band, $17-60$~keV, of
$f_x=5.6\times10^{-11}$~erg~cm$^{-2}$~s$^{-1}$ and luminosity
$L_x=7.1\times10^{43}$~ergs~s$^{-1}$. The corresponding flux and luminosity
in the $2-10$~keV energy band are
$f_x=4.2\times10^{-10}$~erg~cm$^{-2}$~s$^{-1}$ and
$L_x=5.1\times10^{44}$~ergs~s$^{-1}$. The flux and luminosity uncertainties
in these bands are dominated by systematic errors which we estimate to be of
order $5\%$.

%%%%%%%%%%%%%%%%%%%%%%%%%%%%%%%%%%%%%%%%%%%%%%%%%%%%%%%%%%%%%%%%%%%%%
\begin{figure}
\vspace*{-3mm}
\centerline{\includegraphics[width=\columnwidth]{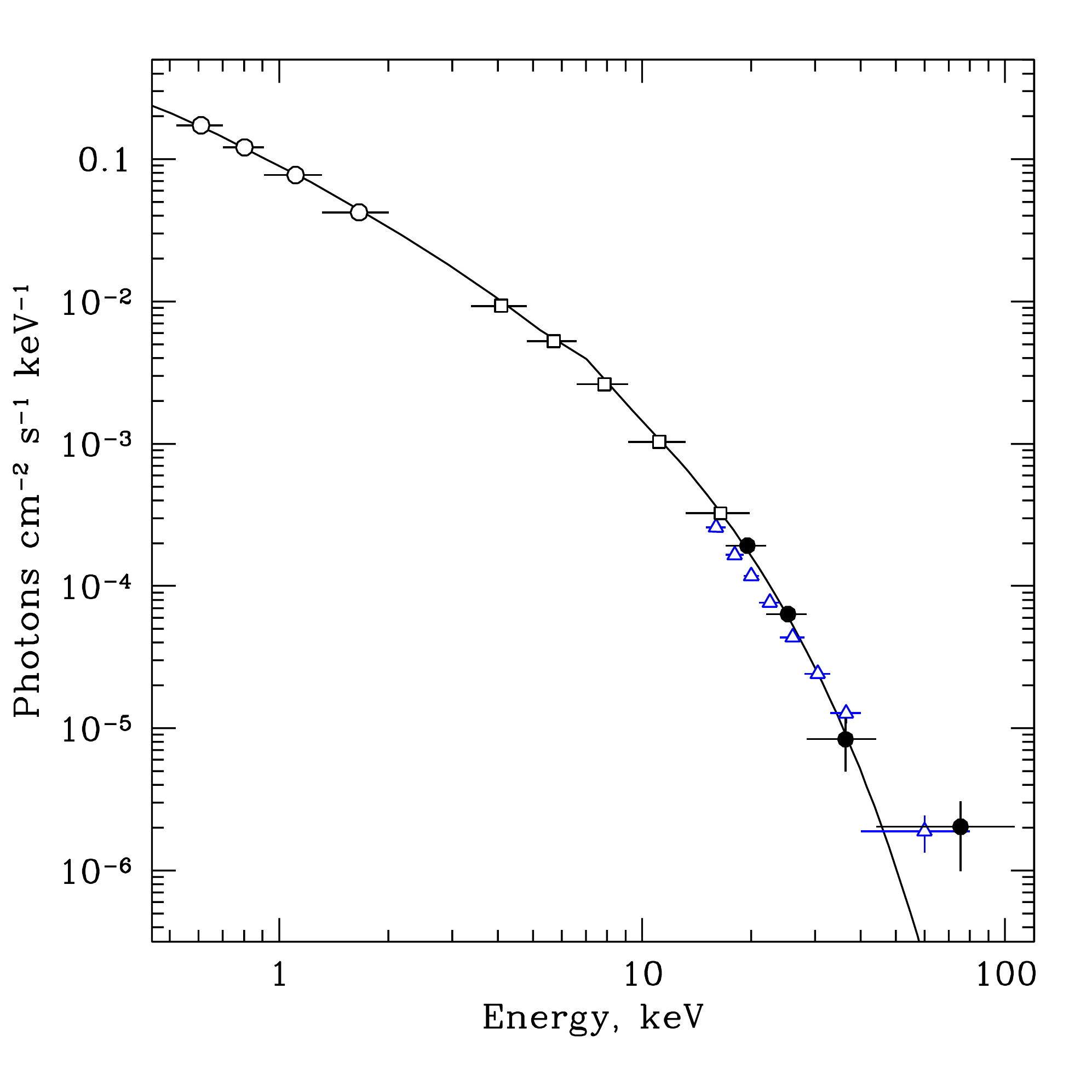}}
\caption{Photon spectra of Coma, derived in this work (symbols
  with the same meaning as in Fig.\ref{spectrum}) and from
  \emph{Beppo-SAX} \citep[open triangles,][]{fus04}. We did not adjust the
  normalization of the \emph{Beppo-SAX} spectrum to account for the
  cluster flux outside the collimator FOV. Note the overall mismatch in
  spectral slope between our \emph{RXTE} and \emph{INTEGRAL}
  measurements in the 15--50~keV energy band. This has implications for
  the decomposition of the observed data into thermal and non-thermal
  components.}
  \label{spectrum:phot}
\end{figure}
%%%%%%%%%%%%%%%%%%%%%%%%%%%%%%%%%%%%%%%%%%%%%%%%%%%%%%%%%%%%%%%%%%%%%

\emph{INTEGRAL} fluxes at $E\lesssim44$~keV are in full agreement with
the thermal spectrum. The thermal component (including possible
contribution from hotter regions within the cluster, see below) should
make a negligible contribution at higher energies. The
\emph{INTEGRAL}-measured flux in the 44--107~keV band is
$(1.8\pm1.1)\times 10^{-11}$~ergs~cm$^{-2}$~s$^{-1}$. If real, this
emission cannot be attributed to the high energy tail of the thermal
spectrum and should instead represent a non-thermal component. We note,
however, that the statistical significance is low ($1.6\,\sigma$) and
\emph{INTEGRAL} statistical uncertainties are comparable to, or higher than,
those in the earlier \emph{RXTE} or \emph{Beppo-SAX} measurements
\citep{rep99,rep02,fus99,fus04}. We also tried to repeat the
\citet{fus04} procedure by fixing the photon index of a non-thermal
component at $\Gamma=2$ and fitting the total spectrum with a thermal
plus power law model. This gives a power law flux of
$(6.0\pm8.9)\times10^{-12}$~ergs~cm$^{-2}$~s$^{-1}$ in the $20-80$~keV
energy band \citep[same as in][]{fus04}. This is a factor of 2.5 lower
than the flux reported by \citet{fus04} in the same energy band even
though we \emph{INTEGRAL} and \emph{Beppo-SAX} appear to measure the
same flux above $\sim 50$~keV (Fig.\ref{spectrum:phot}). The source of
discrepancy seems to be instead in the fluxes measured at the lower
boundary of the \emph{Beppo-SAX} band. The \emph{Beppo-SAX} data points
below 20 keV seem to be a factor of 1.5--2 below our \emph{RXTE} and
\emph{INTEGRAL} measurements, and hence they imply a lower thermal
component flux. These discrepancies underscore the importance of using a
broad-band spectra for detection of non-thermal components from hot
clusters at around $E=50$~keV.

Where \emph{INTEGRAL} can do significantly better than the previous
observatories is to check whether this emission can be attributed to
single point source within the cluster. We do not detect any significant
point sources in the hard-band image (see below) and a $2\sigma$ upper
limit on the point source flux is
$7.6\times10^{-12}$~ergs~cm$^{-2}$~s$^{-1}$ in the 44--107~keV band.
This is below the fluxes of non-thermal components reported by
\cite{fus04} and \cite{rep02}, and so we can exclude a possibility that
these detections can be attributed to a single persistent AGN.

\subsection{Imaging analysis}

\begin{figure}[t]

\includegraphics[width=8.5cm,bb=50 155 530 635]{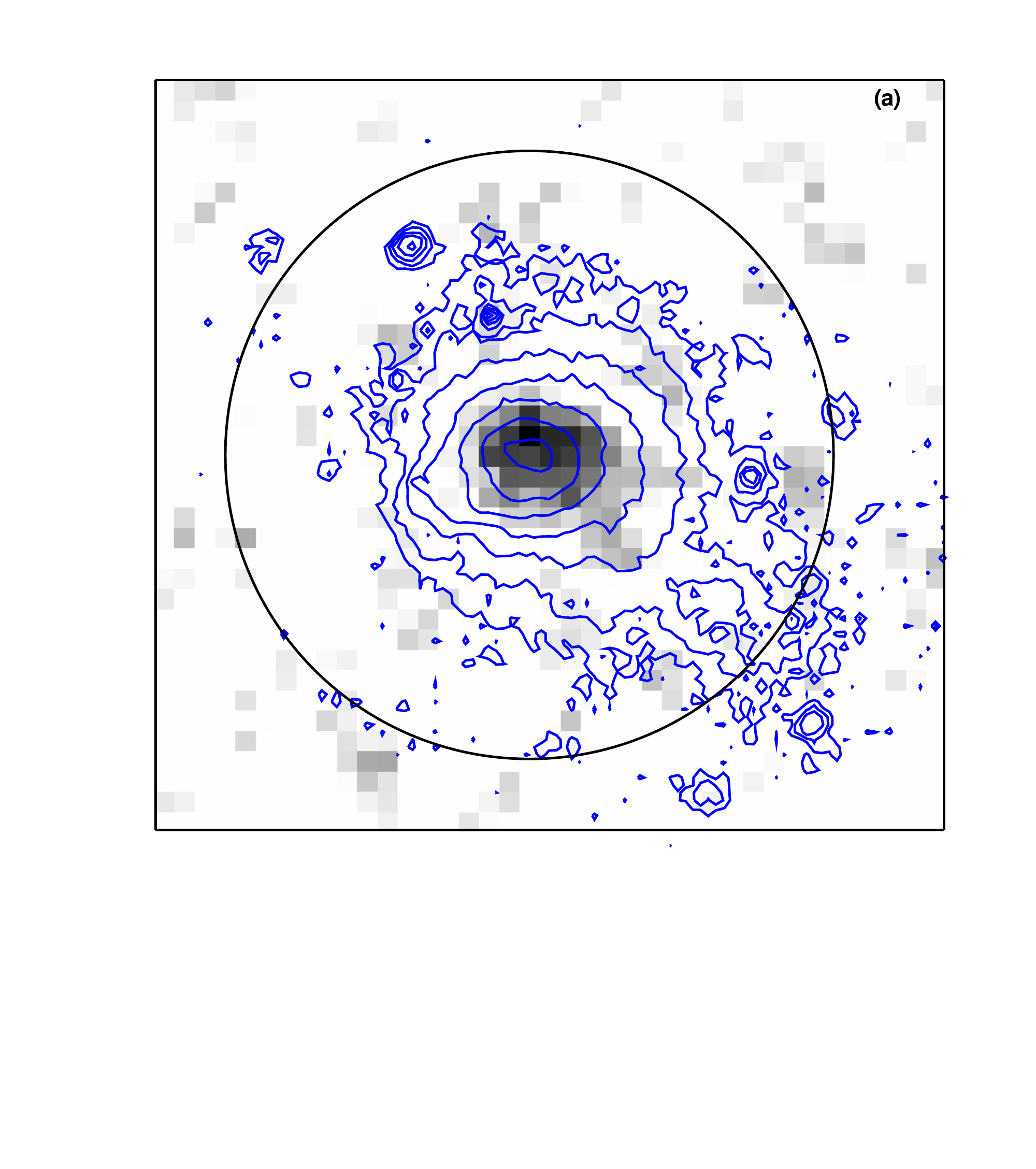}

\includegraphics[width=8.5cm,bb=50 155 530 615]{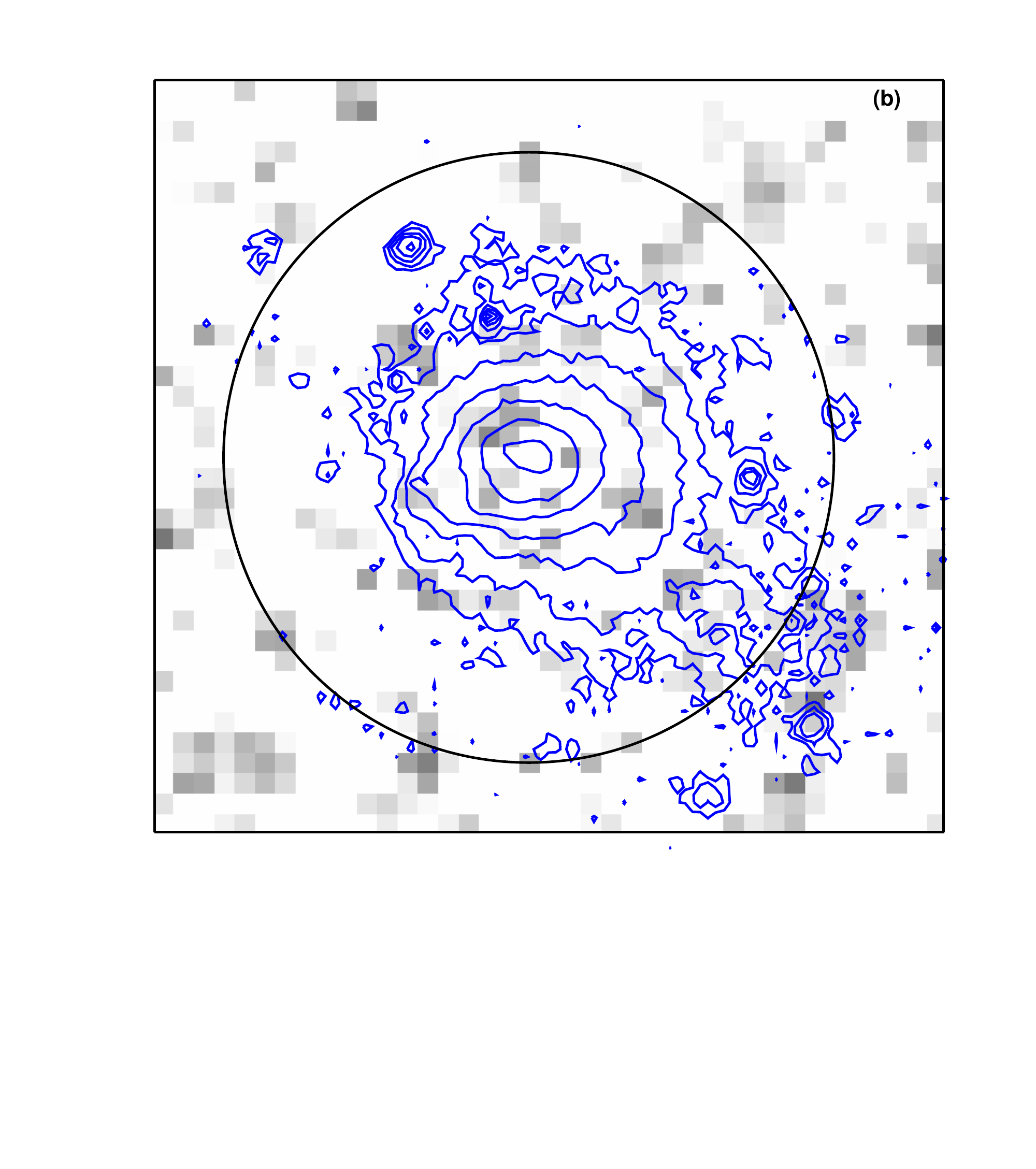}

\caption{\emph{INTEGRAL} images ($\sim3\degr\times3\degr$) of the Coma
  region obtained in $17-28.5$ keV (a) and $44-107$ keV (b) energy ranges.
  Contours represent the \emph{ROSAT} intensity levels in the $0.5-2$ keV energy
  band. Circles indicate an $r=60$\arcmin\ regions around the
  cluster.}\label{image}
\end{figure}

To study the spatial structure of the cluster emission, we used images in
the $17-28.5$~keV (``soft'') and $44-107$~keV (``hard'') \emph{INTEGRAL}
bands. The first band is a combination of two spectral channels where the
cluster emission was detected with high significance; the second band is
where the putative non-thermal component should dominate the thermal plasma
emission. The images are shown in Fig.\ref{image}. In the soft band, Coma is
clearly an extended source \citep[see][for detailed modeling of the
\emph{INTEGRAL} image]{eck07}; a point source would be confined to a
$\sim 3\times3$ pixels square in these images. 

The contours in Fig.\ref{image} show the \emph{ROSAT} surface brightness
levels in the 0.5--2~keV band. The \emph{INTEGRAL} soft-band image shows an
elongation towards the infalling subcluster and a small offset relative to
the \emph{ROSAT} surface brightness peak \citep[see also][]{eck07}. The
off-set is $\sim4.3\arcmin$ to the West (Fig.\ref{image}); it is small but
significant (e.g., the locations of the two AGNs detected in the Coma field,
NGC~4151 and NGC~4388 are within $0.2'$ of their optical positions). As we
discuss below, the offset between \emph{INTEGRAL} and \emph{ROSAT} images
most likely reflects the temperature variations within the cluster.

%%%%%%%%%%%%%%%%%%%%%%%%%%%%%%%%%%%%%%%%%%%%%%%%%%%%%%%%%%%%%%%%%%%%%%%%%
%%%%%%%%%%%%%%%%%%%%%%%%%%%%%%%%%%%%%%%%%%%%%%%%%%%%%%%%%%%%%%%%%%%%%%%%%

%%%%%%%%%%%%%%%%%%%%%%%%%%%%%%%%%%%%%%%%%%%%%%%%%%%%%%%%%%%%%%%%%%%%%%%
\begin{figure}
\includegraphics[width=\columnwidth,bb=50 155 530 625,clip]{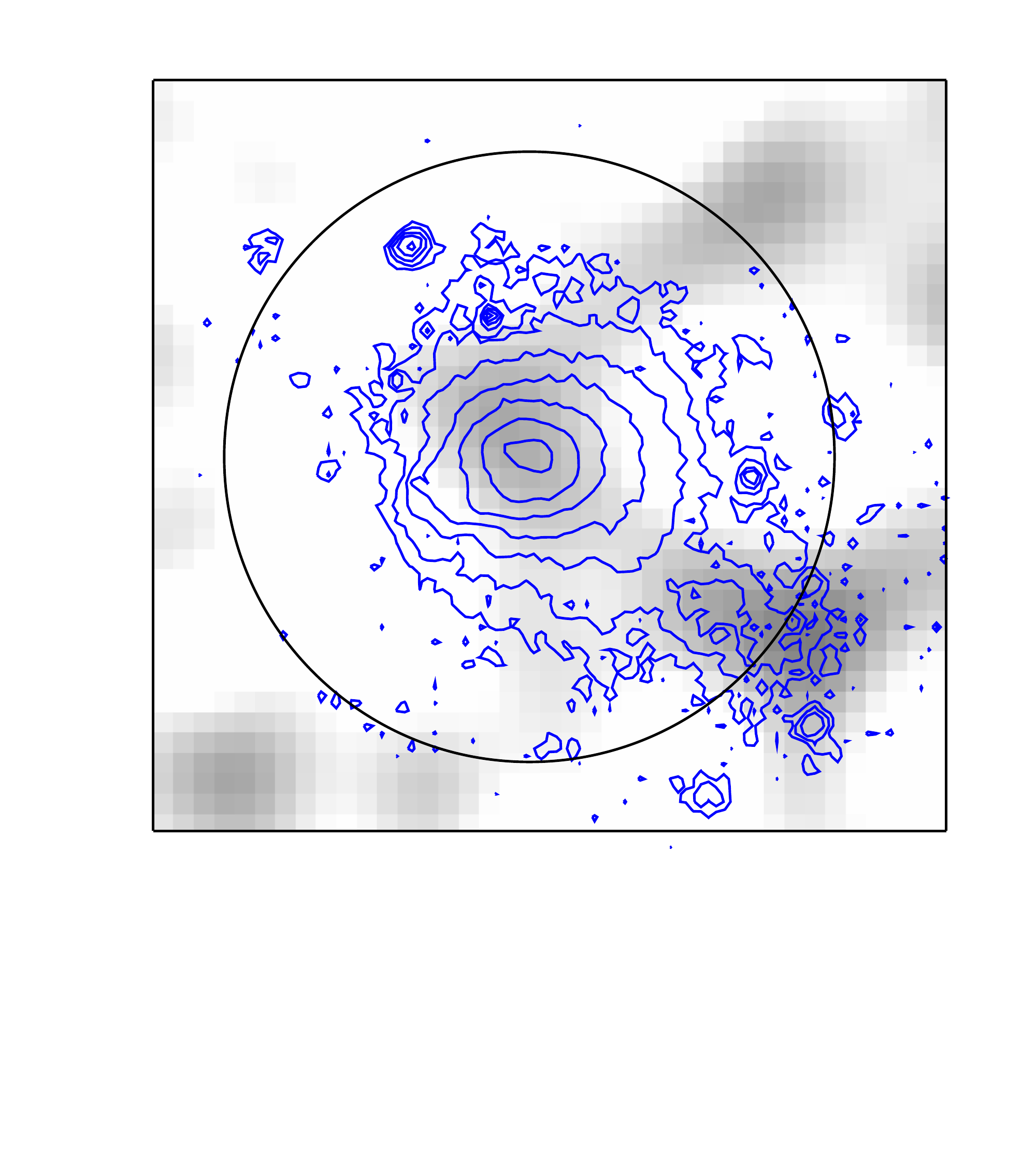}
\caption{Same as in Fig.\ref{image}b, but converted with the beta-model. The gray scale
is different than in Fig.\ref{image}b. \label{ima_smo}}
\end{figure}
%%%%%%%%%%%%%%%%%%%%%%%%%%%%%%%%%%%%%%%%%%%%%%%%%%%%%%%%%%%%%%%%%%%%%%%

We now discuss the hard-band \emph{INTEGRAL} image (44--107~keV). As was
discussed above, the total flux within 60\arcmin{} was detected in this band
with a $1.6\,\sigma$ significance. Can this emission be associated with a
small number of point sources or with an extended component centered on the
cluster? The raw image (Fig.\ref{image}b) does not show any
significant structures. The brightest spots correspond to $1.5-2\,\sigma$
significance. In particular, we can exclude a possibility that the flux
measured by \emph{INTEGRAL} in this band can be attributed to a single point
source; a $2\,\sigma$ upper limit is $\sim7.6\times10^{-12}$ ergs cm$^{-2}$
s$^{-1}$. Figure~\ref{ima_smo} shows the hard-band image smoothed with the
\emph{ROSAT} $\beta$-model to illustrate the distribution of a more extended
component. Again, there are no statistically significant features in the
smoothed image. A local maximum near the Coma center is offset from the
cluster location and in fact is not the brightest feature in the
field. Therefore, we cannot conclude that the hard emission is spatially
coincident with the Coma cluster; it is more consistent with being a
statistical fluctuation.

\subsection{Temperature Map}

\emph{INTEGRAL}'s ability to reconstruct X-ray images in the 20--30~keV band
opens a unique opportunity to compute the temperature map of Coma simply
from the ratio of \emph{INTEGRAL} and \emph{ROSAT} surface brightnesses. The
\emph{INTEGRAL} energy band is above the exponential cutoff in the thermal
$T\sim 10$~keV spectrum, thus the resulting temperature map is very robust
and relatively insensitive to the calibration uncertainties in either
instrument.

The drawback is a relatively poor PSF of \emph{INTEGRAL}/IBIS which
results in a poor spatial resolution in the resulting temperature
map. The PSF angular size is comparable to the cluster core-radius and
so the \emph{ROSAT}-to-\emph{INTEGRAL} ratio should take into account
the redistribution of flux from the center to large radii. To
improve statistics, we smoothed the \emph{INTEGRAL} image
(Fig.\ref{image}a) with a Gaussian with $\sigma=5'$ that approximates
the IBIS PSF in the mosaic images containing observations with
different rotation angles \citep{kri07}. The smoothed image has an
effective PSF that can be approximated by a $\sigma=7.1'$ Gaussian. To
match the resolution of soft- and hard-band data, we smoothed the
\emph{ROSAT} image with the same Gaussian before computing the ratio
map.

%%%%%%%%%%%%%%%%%%%%%%%%%%%%%%%%%%%%%%%%%%%%%%%%%%%%%%%%%%%%%%%%%%%%%
\begin{figure}[t] 
\includegraphics[width=\columnwidth,bb=37 151 577 660]{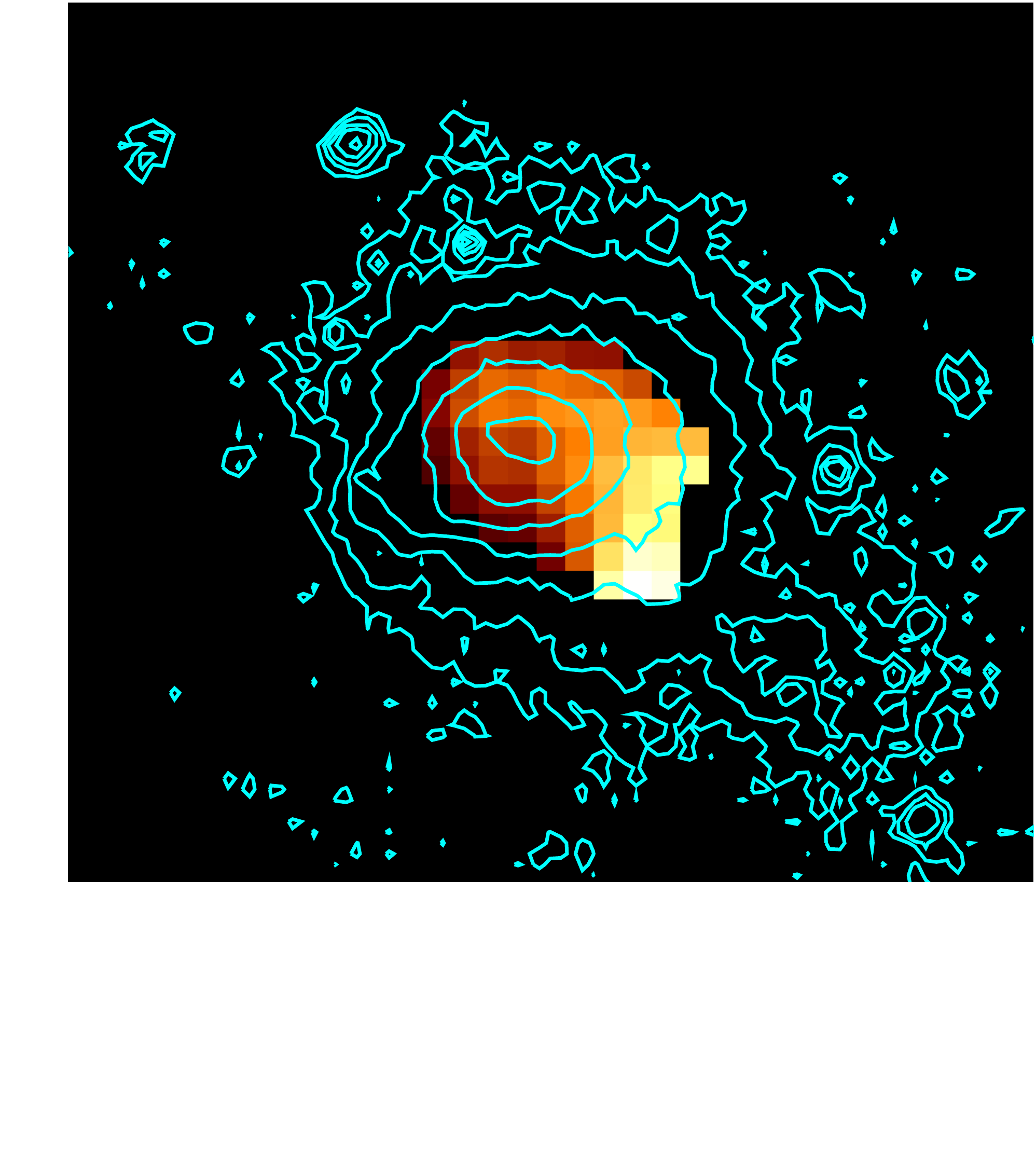}
\caption{The temperature map of the central part of the Coma cluster.
Temperatures of 7.5--8.5 keV, 9.5-10.5 keV and more than 11 keV
correspond to brown, yellow and white colors, respectively. Contours
represent the \emph{ROSAT} intensity levels.}  \label{tempmap}
\end{figure}
%%%%%%%%%%%%%%%%%%%%%%%%%%%%%%%%%%%%%%%%%%%%%%%%%%%%%%%%%%%%%%%%%%%%

The obtained \emph{INTEGRAL}-to-\emph{ROSAT} ratio map was converted to
the temperature map using a lookup table of 17--28.5 to 0.5--2~keV flux
ratios for the MEKAL model as a function of temperature. As can be
expected, this ratio is very sensitive to the temperature (varies by a
factor of $\sim 10$ when $T$ varies from 5 to 12 keV). An approximately
$10\,\sigma$ detection of Coma by \emph{INTEGRAL} translates into
$\sim0.3$ keV temperature uncertainties in the central region. The
resulting temperature map is shown in Fig.\ref{tempmap}. The map is
restricted to the region where the hard X-ray flux is detected with at
least $2.5\,\sigma$ significance.  Qualitatively, our temperature map is
similar to that obtained by \citet{arn01} from the \emph{XMM-Newton}
data. The most notable features are the cold region to the South-East of
the center coinciding with a filamentary structure in the X-ray
brightness \citep{vih97} and the hot region in the direction of the
infalling subcluster\footnote{This hot region was also identified by
  \cite{eck07} with a similar technique (ratio of \emph{INTEGRAL} and
  \emph{XMM-Newton} fluxes); their temperature for this region is
  however higher than ours ($\sim12$ keV vs $\sim9.7$ keV at the
  distance of $\sim14$\arcmin\ from the center to south-west) probably
  because of neglecting the \emph{INTEGRAL} PSF effects}.

\section{Discussion}

The possibility of using radio (synchrotron) and X-ray (Inverse Compton)
observations to constrain the strength of the magnetic fields in plasma has
been extensively discussed in application to various astrophysical sources
(e.g. Felten \& Morrison 1966, Tucker et al., 1973, Rephaeli 1979). We make
similar estimates using the parameters relevant for \emph{INTEGRAL}
observations of the Coma cluster.  For an electron with the Lorentz factor
$\gamma\gg 1$ moving in a uniform magnetic field $B$ at a pitch angle
$\theta$ the synchrotron emission spectrum peaks at the frequency
\begin{eqnarray}
\nu_r\sim 0.29 \frac{3}{4\pi}\frac{eB\sin\theta}{m_ec}\gamma^2
\label{eq:nur}
\end{eqnarray}
\citep[e.g.,][]{gin65}, where $m_e$, $e$, $c$ are the electron mass and
charge, and the speed of light, respectively. The synchrotron emission
of the Coma halo was observed at frequencies ranging from 30 MHz up to
1.4 GHz, and was shown to have spectral index $\alpha\sim 1.34$ (see
e.g.  Kim et al., 1990, Deiss et al., 1997 and references therein).  At
higher frequencies the evidence for spectrum steepening has been
reported (Schlickeiser, Sievers \& Thiemann 1987, Thierbach, Klein, \&
Wielebinski 2003). Below we will use the halo flux $F_r=0.64\pm0.035$ Jy
at $\nu_r=1.4$ GHz reported by \citet{dei97}.  From eq.\ref{eq:nur} it
follows that emission at this frequency is provided by electrons with
$\gamma_r\sim 10^4$ if the field is order of $10~\mu G$.  We further
assume that the surface brightness distribution at 1.4 GHz can also be
approximated by a $\beta$-model;
\begin{eqnarray}
S_r(x)\propto \left[ 1+\left (x/r_{c,r}\right )^2 \right
]^{-3\beta_r+0.5}, \label{eq:sbr}
\end{eqnarray}
where $r_{c,r}$ and $\beta_r$ are the co-radius and beta-parameter of the
radio surface brightness. According to \citet{dei97} the scale size of the
radio halo at 1.4 GHz is similar to that of the X-ray halo parameters
derived by \citet{bri92} (i.e.  $r_{c,r}\sim 10.'5$, but the radio declines
with radius more rapidly (i.e. $\beta_r > \beta_X=0.75$). \citet{col05}
fitted the surface brightness distribution from \citet{dei97} with
$r_{c,r}=23'.8$ and $\beta_r=1.47$. The volume emissivity $\Upsilon(r)$
of the cluster (assuming spherical symmetry) at 1.4 GHz is then
\begin{eqnarray}
\Upsilon_r(r)\propto \left[ 1+\left ( r/r_{c,r}\right )^2
\right ]^{-3\beta_r}. \label{eq:emr}
\end{eqnarray}

\begin{figure}[t]
\includegraphics[width=\columnwidth,bb=0 0 560 570]{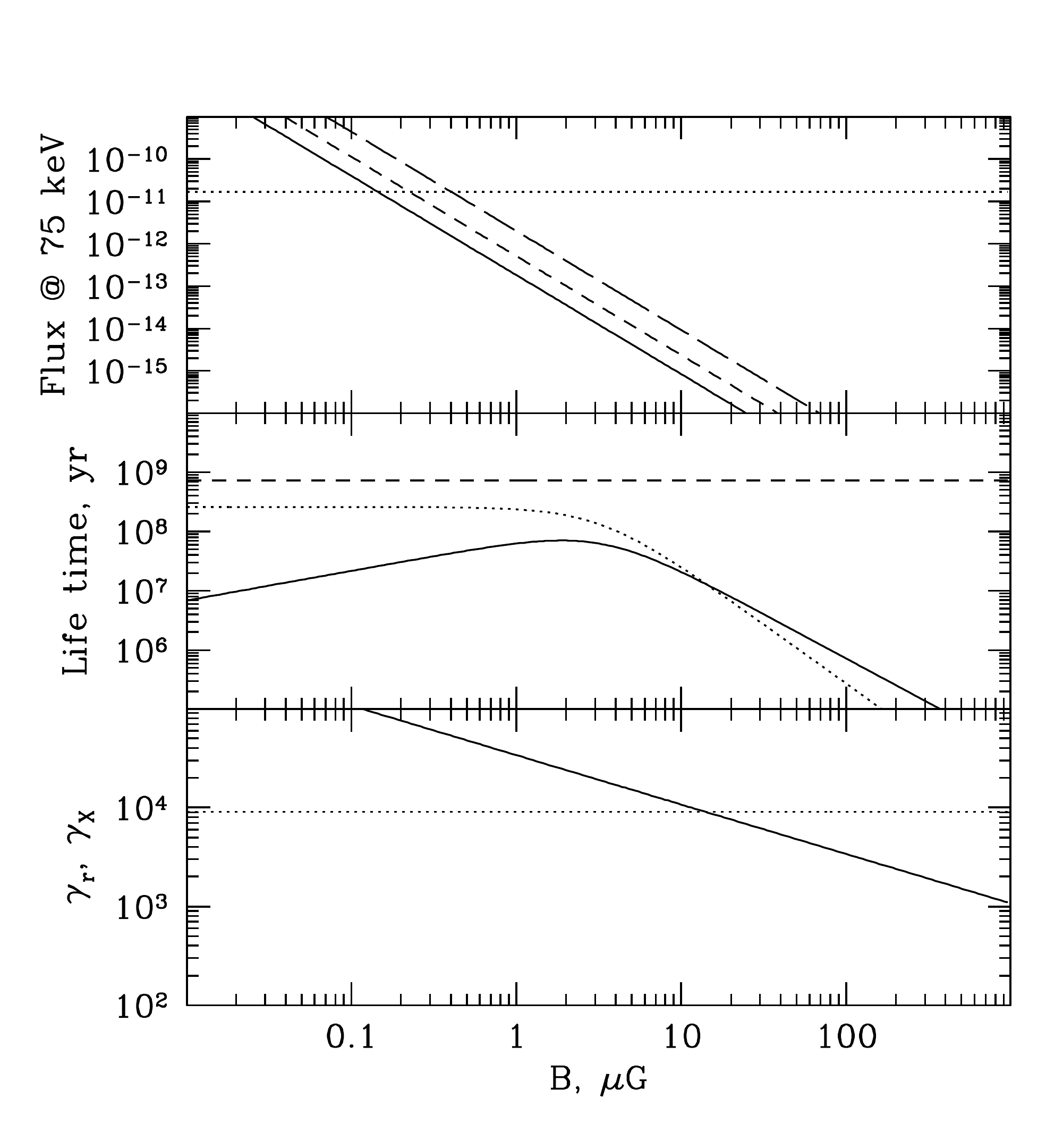}
\caption{Bottom panel: Lorentz factor of electrons emitting synchrotron
  radiation at 1.4 GHz ($\gamma_r$, solid line) as a function of magnetic
  field $B$ and the Lorentz factor of electrons producing 75 keV photons by
  up-scattering CMB radiation ($\gamma_X$, dotted line). Middle panel: The
  life time of electrons having Lorentz factors $\gamma_r$ $\gamma_X$ (solid
  and dotted lines respectively). The top panel shows the comparison of the
  spectral intensity at 75 keV ($\nu F_\nu$, in units
  of~ergs~cm$^{-2}$~s$^{-1}$, horizontal dotted line) corresponding to the
  observed \emph{INTEGRAL} 44--107 flux, $(1.8\pm1.1)\times
  10^{-11}$~ergs~cm$^{-2}$~s$^{-1}$, with the expected level of IC emission.
  The model IC fluxes were computed using eq.\ref{eq:fx} for the halo radio
  flux $F_r=0.64$ Jy at 1.4 GHz and uniform magnetic field $B$. The short
  and long dashed lines show the same IC level, but scaled by factors of 2.8
  and 11 to illustrate the changes introduced by the assumption that
  magnetic fields energy density declines with radius according to
  eq.\ref{eq:bscale} with the parameter $a=0.5$ and $a=1$ respectively. In
  this case the value of $B$ shown along X axis is the value of magnetic
  field $B_0$ at the center of the cluster.}
\label{fig:comab}
\end{figure}

We now can estimate the hard X-ray flux due to Inverse Compton scattering of
CMB photons by relativistic electrons.  The electron with the Lorentz factor
$\gamma\gg 1$ will up-scatter CMB photons to a characteristic frequency
\begin{eqnarray}
\nu_x\sim\frac{4}{3}\gamma^2\nu_{CMB}\sim\frac{4}{3}\gamma^2 \frac{3\,
kT_{CMB}}{h}, \label{eq:nux}
\end{eqnarray}
where $T_{CMB}\approx 2.7$K is the CMB temperature and $k$ is the Boltzmann
constant. Thus for the 75 keV photon the required electron Lorentz factor is
$\gamma_X\sim9\times10^3$, i.e. close to the value needed to produce radio
emission at 1.4 GHz for plausible values of magnetic fields (see
Fig.\ref{fig:comab}, bottom panel). The expected IC flux due to CMB photons
up-scattering produced by the electrons responsible for synchrotron emission
is \citep{fel66}:
\begin{eqnarray}
F_x=F_r\times C(p) \left( \frac{\nu_r}{\nu_x}\right)^\alpha
T_{CMB}^{\alpha+3} B^{-\alpha-1}, \label{eq:fx}
\end{eqnarray}
where $C(p)$ is a function of the power law slope of the electron energy
spectrum, $p$, which is related to the synchrotron spectral index as
$p=2\alpha+1$.

%%%%%%%%%%%%%%%%%%%%%%%%%%%%%%%%%%%%%%%%%%%%%%%%%%%%%%%%%%%%%%%%%%%%%
%%%%%%%%%%%%%%%%%%%%%%%%%%%%%%%%%%%%%%%%%%%%%%%%%%%%%%%%%%%%%%%%%%%%%

%%%%%%%%%%%%%%%%%%%%%%%%%%%%%%%%%%%%%%%%%%%%%%%%%%%%%%%%%%%%%%%%%%%%%
\begin{figure}[t]
\includegraphics[width=\columnwidth,bb=0 0 560 570,clip]{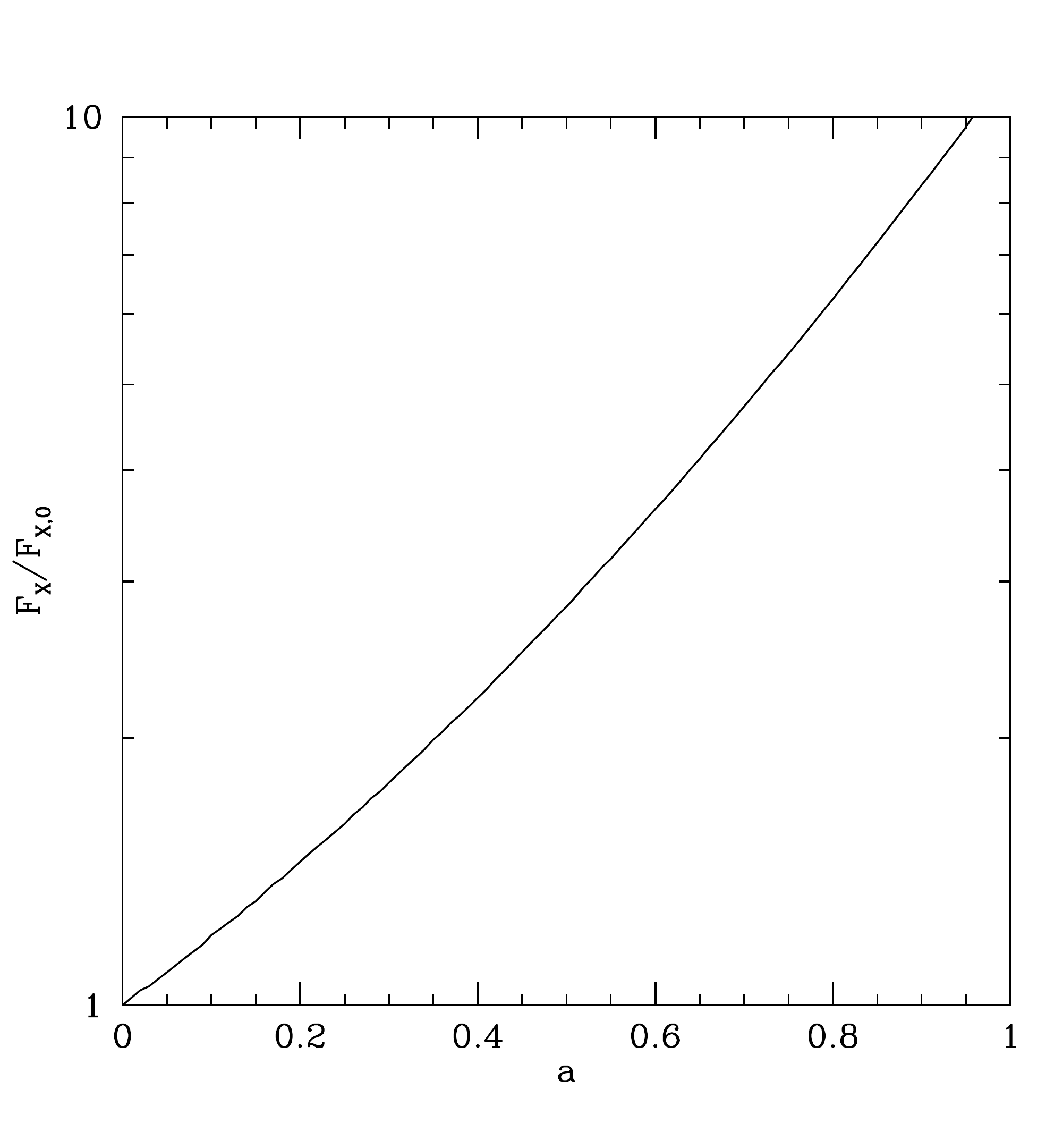}
\caption{Variations of IC flux on the different behavior of the
magnetic field with radius. The parameterization is done in the form
$B^2(r)= B^2_0 \left ( \frac{\rho(r)}{\rho_0}\right )^a$. See text
for details. \label{fig:comar} }
\end{figure}
%%%%%%%%%%%%%%%%%%%%%%%%%%%%%%%%%%%%%%%%%%%%%%%%%%%%%%%%%%%%%%%%%%%%%

Assuming that $B$ is constant across the cluster, one can calculate
expected IC flux for a given $B$ and radio flux $F_r$ using
eq.\ref{eq:fx} as is done in the top panel of Fig.\ref{fig:comab}. The
intersection of the predicted (solid line) and observed (dotted line)
curves gives the value of the magnetic field $\sim$ 0.1-0.2~$\mu$G
needed to explain radio and hard X-ray fluxes as the synchrotron and
IC emission respectively produced by the same electron population.

The assumption of constant $B$ across the cluster probably is too
simplistic, especially given that energy density of the ICM varies strongly
from the center to outskirts (see, e.g., Dolag, Bartelmann \& Lesch, 2002).
The spatial variations of the magnetic field would affect the relation of
the synchrotron and IC fluxes (Rephaeli 1979; Goldshmidt \& Rephaeli 1993;
Brunetti et al.\ 2001; Newman et al.\ 2002; Colafrancesco et al.\ 2005). We
estimate the effects of magnetic field variations using a simple
parameterization,
\begin{eqnarray}
  B^2(r)= B^2_0 \,(\rho(r)/\rho_0)^a,\label{eq:bscale}
\end{eqnarray}
where $\rho(r)$ is the thermal gas density, which scales approximately as
the ICM pressure if the gas density variations are much stronger than those
of temperature over the region of interest. Adopting the \emph{ROSAT}
$\beta$-model for the gas density distribution, we can write:
\begin{eqnarray}
B(r)= B_0 \left[ 1+\left ( x/r_{c,x}\right )^2 \right
]^{-\frac{3}{4}a\beta_x},
\end{eqnarray}
where $B_0$ is the magnetic field at the center of the cluster. In this
parameterization $a=0$ corresponds to the case of a constant magnetic field,
while $a=1$ implies that energy density of the magnetic field scales as
the ICM pressure. Compared to the constant $B$ case (i.e. eq.\ref{eq:fx})
the ratio of the IC and synchrotron fluxes (within given distance from the
center) has to be multiplied by an additional factor $f(a)$:
\begin{eqnarray}
f(a)=\frac{\int{\Upsilon_r(r) \left (
\frac{B}{B_0}\right)^{-\alpha-1} r^2dr}}{\int \Upsilon_r(r)r^2dr},\label{eq:fa}
\end{eqnarray}
where $\Upsilon_r(r)$ is given by eq.\ref{eq:emr} where the integration is
in the range $r=0-60'$. The computed factor $f(a)$ is shown in
Fig.\ref{fig:comar} (in these calculations we assumed the X-ray surface
brightness parameters $r_{c,r}=r_{c,x}=10'.68$, $\beta_r=\beta_x=0.741$).

The total IC flux (up to infinity) may actually diverge for large $a$ since
the drop of the $B$ at large distances from the center has to be
``compensated'' by energy density of relativistic particles to maintain the
assumed $\Upsilon{E}(r)$. This causes strong increase of IC
emission. Formally for convergence (up to infinity) one needs:
\begin{eqnarray}
a<\frac{2(2\beta_r-1)}{\beta_x(\alpha+1)},
\end{eqnarray}
which translates to $a<0.55$ for our parameters. However, since we integrate
the hard X-ray flux only within a 1\degr\ radius, we can use wider range of
$a$. It follows from Fig.\ref{fig:comar} that the change in $a$ from 0 to 1
changes the ratio $f(a)$ by a factor of $\sim$11.  This translates into the
uncertainty in the estimated $B_0$ by a factor of $f(a)^{1/(1+\alpha)}\sim
2.8$. An additional uncertainty is introduced by our determination of the
hard X-ray flux which was done by convolution of the \emph{INTEGRAL} image
with the surface brightness model observed at $E=1$~keV by \emph{ROSAT}.
If, for example, the true surface brightness in the hard band has the same
core-radius but $\beta$ is varied in the range 0.3-1.5, this would change
our derived values of $B_0$ by $\pm$30\%.

Although the modeling uncertainties described above are substantial, they do
not change the conclusion that it is unlikely that the IC emission in the
$44-107$ keV band can reach the flux levels comparable to the
ISGRI/\emph{INTEGRAL} sensitivity, unless the magnetic field is as low as a
few$\times0.1~\mu G$. For these low values of the magnetic field the life
time is shorter for electrons responsible for the radio synchrotron emission
(Fig.\ref{fig:comab}) and thus the IC spectrum of the electron population
has to be close to a power law, as long as a power law spectrum is observed
in the radio.

If we treat the observed hard X-ray flux as a detection, the derived $B\sim
0.1-0.2$~$\mu$G is an order of magnitude below the values derived from
Faraday rotation $B\sim 1.7$~$\mu$G \citep[e.g.,][]{kim90}. Various ways to
bring these estimates into agreement have been discussed (see e.g.
Goldshmidt \& Rephaeli 1993, Petrosian 2001, Brunetti et al., 2001, Newman
et al. 2002).  The most obvious factor affecting these estimates is the
spatial variations of the field strength.  Note that the Faraday rotation
signal depends on the line of sight integral of the product of the electron
density and the parallel component of the field. The value of the field
derived from the comparison of the cluster IC and radio fluxes depends
instead on the volume averaged non-linear function of the field (see
eq.\ref{eq:fa}). Thus different assumptions on the non-uniformity of the
magnetic field will affect differently the estimates of $B$ with these two
methods and the above mentioned discrepancy can be at least partly removed.

However, given the low significance of the signal, we should treat the
\emph{INTEGRAL} results as an upper limit on the IC flux. Our $2\sigma$
upper limit in the $44-107$ keV band is perfectly consistent with the
magnetic fields stronger than few$\times0.1~\mu G$. For even larger fields,
the expected IC signal drops as $\approx B^{-\alpha-1}\approx B^{-2.34}$ and
will be below the reach of even planned future experiments. For example, for
$B=10\,\mu$G, the expected IC flux at $E=75$~keV is a factor of $\sim 10^4$
below the \emph{INTEGRAL} sensitivity, even without account for any possible
electron aging (see Fig.\ref{fig:comab}).

\section{Summary}

The summary of main results from the deep observations of the
\emph{Coma} cluster with \emph{INTEGRAL} and combining these
observations with results of \emph{RXTE} and {ROSAT} observatories lead
us to the following conclusions:

1) The total cluster spectrum in the 0.5--50~keV energy band is well
described by the thermal plasma emission with the mean temperature
$T\simeq8.2$ keV;

2) There are significant temperature variations within the cluster and the
mean 8.2~keV temperature is the result of mixing emission components with
$T$ ranging from 7.5 to 10.5~keV. The temperature near the
\emph{ROSAT} surface brightness peak is $\sim8.5$~keV.

3) We do not detect a significant excess over the thermal spectrum at high
energies. The upper limit on the non-thermal flux is, however, consistent
with the previous detections reported on the basis of \emph{Beppo-SAX} and
\emph{RXTE} observations.

4) It is unlikely that IC emission in hard X-rays can reach the flux level
comparable with the \emph{INTEGRAL}/ISGRI sensitivity for the magnetic
fields strongerr than few$\times0.1~\mu$G. A $2\sigma$ upper limit of the
hard X-ray flux from \emph{INTEGRAL} is consistent with $B\lesssim1\,\mu$G.

\acknowledgments

This work is based on observations with \emph{INTEGRAL}, and made use of
the \emph{INTEGRAL} Science Data Center (Versoix, Switzerland), Russian
Science Data Center of \emph{INTEGRAL} (Moscow, Russia) and the High
Energy Astrophysics Science Archive Research Center Online Service,
provided by the NASA/Goddard Space Flight Center. The work was supported
by the \emph{Chandra} grant GO5-6121A, RFBR grant 07-02-01051, DFG grant
CH389/3-2, the ``Extended objects in the Universe'' program of the
Russian Academy of Sciences, and a program of support for leading
scientific schools (Project NSH-1100.2006.2). AL thanks CfA for
hospitality during the course of this research.

\end{document}